\documentclass[letterpaper,10pt]{article} 

\usepackage{opticameet3} 

\newcommand\authormark[1]{\textsuperscript{#1}}

\usepackage{amsmath,amssymb}
\usepackage[colorlinks=true,bookmarks=false,citecolor=black,urlcolor=black, linkcolor=black]{hyperref} 

\begin{document}

\title{Modulation-Free Laser Stabilization with Extended Locking Range on a SiN Chip}


\author{Mohamad Hossein Idjadi\authormark{1,*}, Farshid Ashtiani\authormark{1}, and Kwangwoong Kim\authormark{1}}

\address{\authormark{1} Nokia Bell Labs, 600 Mountain Ave, Murray Hill, NJ 07974, USA.}

\email{\authormark{*}e-mail: mohamad.idjadi@nokia-bell-labs.com} 

\begin{abstract}
We demonstrate a modulation-free laser stabilization system using a cavity-coupled MZI with aided acquisition on a low-loss SiN chip, achieving more than an order-of-magnitude improvement in locking range and over 36 dB noise suppression.
\end{abstract}

\section{Introduction}
Low phase noise lasers have broad applications, including optical fiber sensing \cite{idjadi2024hybrid, morton2018high}, low phase noise microwave synthesis \cite{kudelin2024photonic}, and coherent optical communication \cite{zhang2024high}. Among various stabilization techniques, electro-optic (EO) frequency locking stands out for its long-term stability. The key component in EO locking circuits is the optical frequency noise discriminator (OFND), which compares the laser frequency to a reference ($f_{ref}$), e.g. the resonance frequency of a microring resonator (MRR), and generates an error signal based on the frequency error.

One approach to implementing an Optical Frequency Noise Detector (OFND) is a Mach-Zehnder Interferometer (MZI) with phase biasing at quadrature, terminated by balanced photodetectors (BPD). While simple, this design offers lower sensitivity (OFND gain) compared to the cavity-coupled MZI (CCMZI) architecture. A modulation-free laser locking system using a CCMZI, as introduced in \cite{idjadi2024modulation, idjadi20246}, leverages cavity resonance to significantly boost sensitivity. However, despite the higher gain offered by the CCMZI, its error signal rapidly diminishes off-resonance, which limits the locking range. Additionally, increasing the Q-factor of the optical reference improves noise suppression but further reduces the locking range as the resonance linewidth narrows. In contrast, the MZI-based OFND maintains a broader locking range, albeit at the cost of reduced OFND gain. This trade-off restricts the use of high-Q cavities in practical applications, such as optical fiber sensing, where locking a laser to a high-finesse and stable cavity is essential. In this work, we propose and experimentally demonstrate a method to overcome this trade-off.

\section{Principle of operation}

\begin{figure}[b]
\centering
\includegraphics[width=\linewidth]{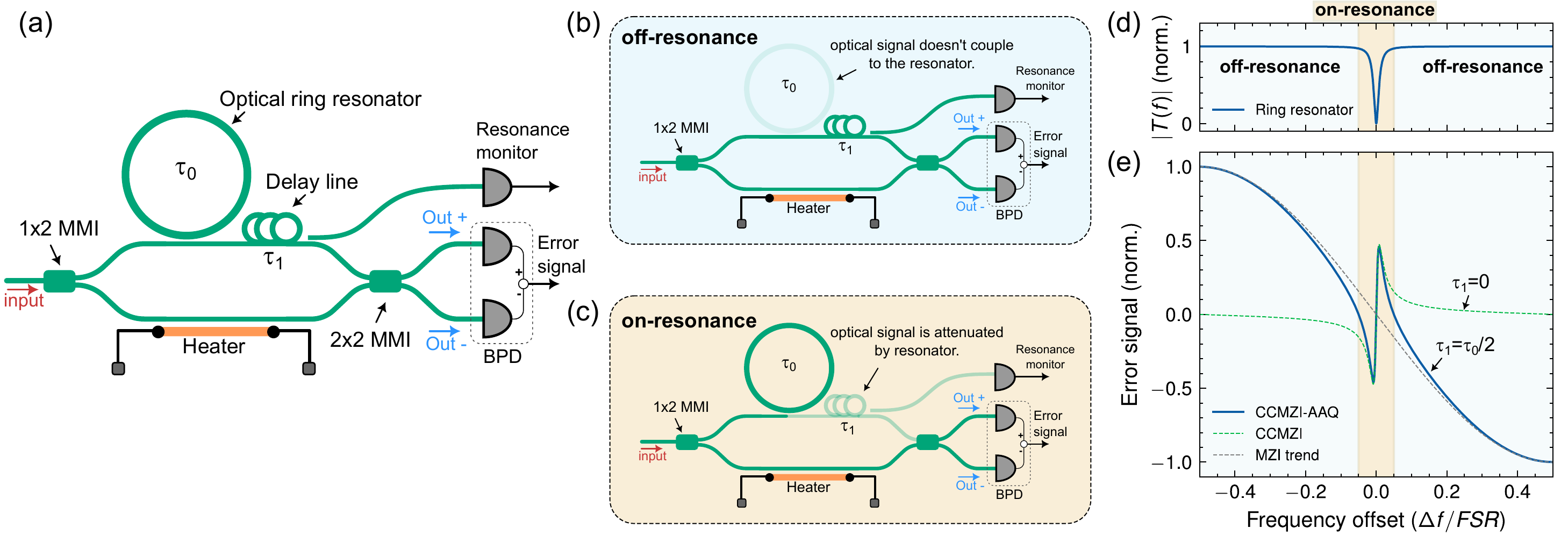}
\caption{\small \textbf{(a)} Schematic of the cavity-coupled MZI with aided acquisition (CCMZI-AAQ). \textbf{(b)} simplified CCMZI-AAQ when the laser is frequency is \textbf{(b)} off-resonance and \textbf{(c)} on-resonance. \textbf{(d)} microring resonator (MRR) transfer function, $|T(f)|$.  \textbf{(e)} The error signal for CCMZI ($\tau_1=0$) and CCMZI-AAQ ($\tau_1=\tau_0/2$).}
\label{fig1_concept}
\end{figure}

Figure \ref{fig1_concept}(a) illustrates the proposed modulation-free laser frequency locking scheme with aided acquisition (CCMZI-AAQ) to enable extended locking range. The input light is split into two branches using a 50\% power splitter (e.g., an MMI). The bottom branch includes a thermal phase shifter for optical phase tuning and to reconfigure the photonic integrated circuit (PIC) across a wide wavelength range, while the top branch uses an integrated MRR as the frequency reference with a free spectral range (FSR) corresponding to time delay of $\tau_0$. After the MRR, an optical delay line with time delay $\tau_1$ is added, with a portion tapped out to monitor the resonance. The two branches then interfere at a 2$\times$2 MMI, directed to BPDs.
Figure \ref{fig1_concept}(b) shows that, off-resonance, the CCMZI-AAQ behaves like a conventional MZI, as minimal signal couples into the ring. Figure \ref{fig1_concept}(c) illustrates that on-resonance, most of the light in the upper branch couples into the ring, minimizing the delay line’s effect. Figures \ref{fig1_concept}(d) and \ref{fig1_concept}(e) show the normalized transfer function of the ring resonator ($|T(f)|$) and the error signal. When $\tau_1=0$, the error signal resembles the conventional CCMZI \cite{idjadi2024modulation}. However, with $\tau_1 = \tau_0/2$, the off-resonance signal resembles a conventional MZI transfer function over one FSR that is used to guide the laser toward $f_{ref}$. The MRR monitor port acts as an on/off-resonance flag signal and once in resonance, the error signal is used in a PID loop to lock the laser frequency to $f_{ref}$. The FSR of the MZI is half that of the MRR, enabling the CCMZI-AAQ to extend the locking range to $(f_{ref} - \Delta/2 , f_{ref} + \Delta/2)$, where $\Delta$ is the FSR of the MRR.

\section{Experimental results}
\subsection{Open loop characterization}

\begin{figure}[b]
\centering
\includegraphics[width=\linewidth]{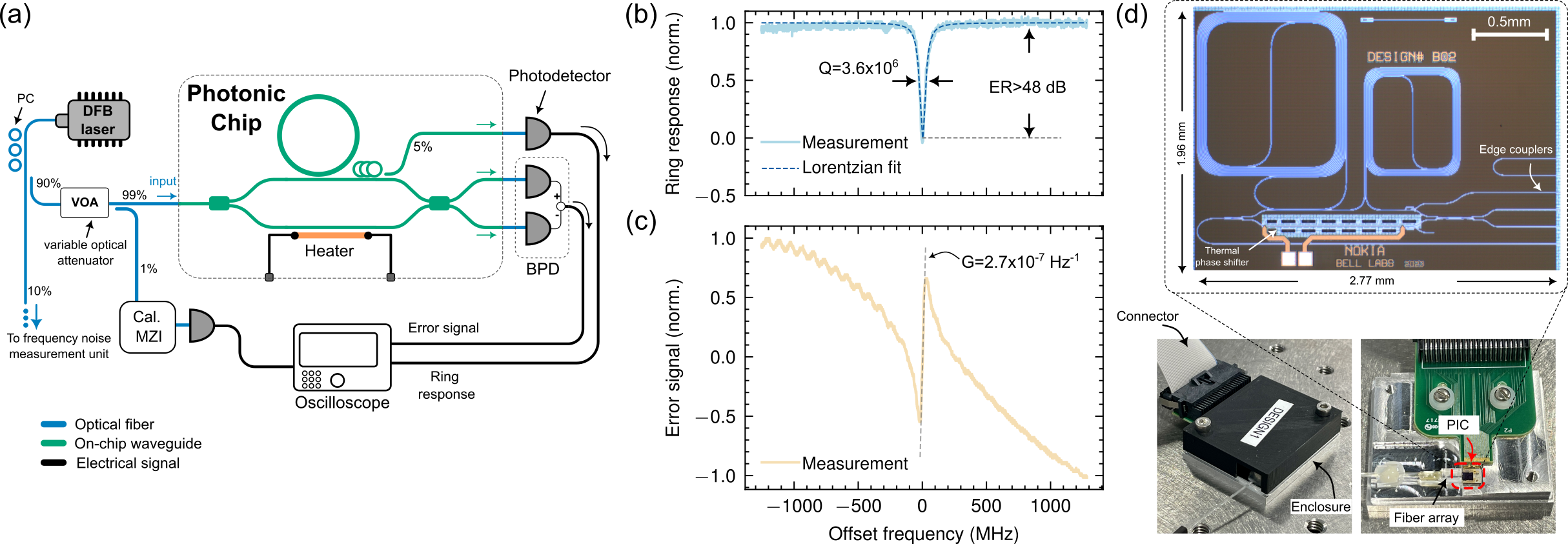}
\caption{\small \textbf{(a)} The PIC characterization and open-loop measurement setup. \textbf{(b)} Measured ring resonator response. \textbf{(c)} measured CCMZI-AAQ error signal. \textbf{(d)} the packaged PIC and the microphotograph of the SiN chip.}
\label{fig2_openloop}
\end{figure}

As a proof of concept, the architecture of the CCMZI-AAQ is implemented on a commercial low-loss silicon nitride (SiN) chip. Figure \ref{fig2_openloop}(a) shows the measurement setup used for characterizing the PIC and the open-loop response. In this setup, the frequency of a distributed feedback (DFB) laser at nominal wavelength of 1552.03 nm is swept by ramping its current. The input and output of the chip are coupled to a UHNA-7 fiber array with a pitch of 127 $\mu$m. The optical signal is coupled into the chip, and both the MRR resonance and balanced outputs are photodetected and monitored using a real-time oscilloscope. Figure \ref{fig2_openloop}(b) shows the measured normalized transfer function of the MRR, with a loaded Q-factor of $3.6\times 10^6$, corresponding to a waveguide loss of approximately 5 dB/m. Figure \ref{fig2_openloop}(c) presents the normalized CCMZI-AAQ error signal, where the OFND gain is approximately $2.7\times 10^{-7}$ Hz$^{-1}$, and the trend aligns well with the simulation results shown in Figure \ref{fig1_concept}(e). Figure \ref{fig2_openloop}(d) displays a microphotograph of the chip, which was implemented using the LIGENTEC AN800 low-loss SiN process \cite{idjadi20246}. The chip occupies an area of $5.43 \ mm^2$ and is packaged with a fiber array and a TEC within an enclosure.

\subsection{Closed-loop characterization}

\begin{figure}[tb]
\centering
\includegraphics[width=\linewidth]{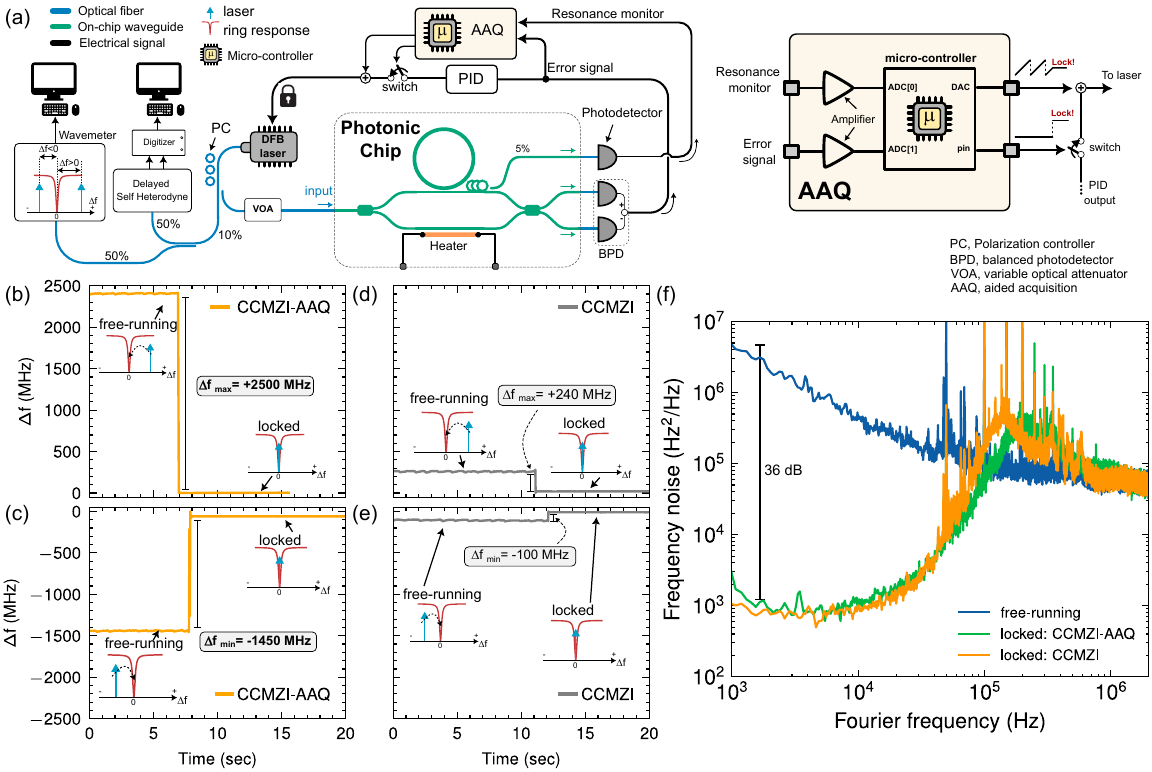}
\caption{\small \textbf{(a)} Closed-loop measurement setup. \textbf{(b, c)} Locking range characterization with the CCMZI-AAQ engaged, showing the laser wavelength measured by a wavemeter when the laser frequency is \textbf{(b)} above and \textbf{(c)} below $f_{ref}$. The maximum measured locking range is 3.95 GHz. \textbf{(d, e)} Frequency locking range measurements with the AAQ disengaged, where the range is reduced to 340 MHz. \textbf{(f)} laser frequency noise measurement. (b,d) and (c,e) share y-axis, while (b,c) and (d,e) share x-axis.}
\label{fig3_closedloop}
\end{figure}

To test the CCMZI-AAQ under closed-loop operation, two measurements were conducted (Fig. \ref{fig3_closedloop}(a)). First, the free-running laser is biased as its maximum positive (Figs. \ref{fig3_closedloop}(b) and \ref{fig3_closedloop}(d) where $\Delta f>0$) and negative (Figs. \ref{fig3_closedloop}(c) and \ref{fig3_closedloop}(e) where $\Delta f<0$) off-resonance offsets at which the loop could lock the laser. In the AAQ system, the MRR resonance is monitored (on-resonance = LOW, off-resonance = HIGH). A micro-controller with a 12-bit DAC and ADC reads both the error signal and the resonance flag signal. If the laser is off-resonance, the micro-controller tunes the laser frequency by ramping its bias current until the resonance monitor signal goes LOW. Once resonance is achieved, the laser is biased at this new current, and the PID controller locks the laser to the MRR resonance frequency, $f_{ref}$, by leveraging the high OFND gain in the on-resonance region. As shown in Figs. \ref{fig3_closedloop}(b) and \ref{fig3_closedloop}(c), engaging the AAQ system extends the locking range to 3.95 GHz, corresponding to the MRR’s FSR. In contrast, as shown in Figs. \ref{fig3_closedloop}(d) and \ref{fig3_closedloop}(e), using only a PID controller without the AAQ system can only lock the laser within a more limited range of approximately 340 MHz. It is worth noting that adjusting the relative phase between the MRR and the MZI can result in a more symmetric locking range.

In the second test, to characterize the effect of additional AAQ on laser frequency noise, the laser frequency noise was measured using a delayed self-heterodyne setup, both with and without the AAQ system. As shown in Fig. \ref{fig3_closedloop}(f), the laser frequency noise is suppressed by approximately 36 dB, and the AAQ system does not degrade the laser noise performance under closed-loop conditions, while significantly extending the locking range.
\section{Conclusion}
A CCMZI is an effective technique for suppressing laser frequency noise while maintaining a simple architecture. However, there is an inherent trade-off between the cavity’s quality factor and the frequency locking range. By incorporating a carefully designed integrated delay line, this trade-off can be mitigated with an AAQ system. The proposed CCMZI-AAQ implemented on a SiN chip achieves over 36 dB of frequency noise suppression and more than an order-of-magnitude improvement in the frequency locking range while occupying 5.43 mm$^2$. This work paves the way for scalable, low-noise lasers with excellent long-term stability, ideal for fiber sensing applications.

\end{document}